\documentstyle[aps,twocolumn,pra,epsf]{revtex}

\begin{document}

\draft

\wideabs{
\title{Atomic Deuterium
Adsorbed on the Surface of Liquid Helium}
\author{A.P. Mosk}
\address{Max-Planck-Institut f\"{u}r
Kernphysik, Saupfercheckweg 1, D-69117, Heidelberg, Germany.}
\author{M.W. Reynolds}
\address{Creo Products Inc., 3700 Gilmore Way, Burnaby, BC,
Canada V5G 4M1.}
\author{and T.W. Hijmans}
\address{Van der Waals-Zeeman Instituut, Valckenierstraat 65,
1018 XE, Amsterdam, Netherlands.}

\date{\today}
\maketitle

\begin{abstract}
We investigate deuterium atoms adsorbed on the surface of liquid
helium in equilibrium with a vapor of atoms of the same species.
These atoms are studied by a sensitive optical method based on
spectroscopy at a wavelength of 122 nm, exciting the 1S-2P
transition. We present a direct measurement of the adsorption
energy of deuterium atoms on helium and show evidence for the
existence of resonantly enhanced recombination of atoms residing
on the surface to molecules.
\end{abstract}


\vspace{4mm}
 }

Triggered by the first observation of Bose-Einstein condensation
(BEC) in atomic gases \cite{cor95,ket95}, there has been a growing
interest in the investigation of quantum degenerate behaviour of
fermionic atoms. Attention has focused mainly on magnetically
trapped gases of the fermionic isotopes $^6$Li and $^{40}$K. The
latter isotope was the first atomic Fermi gas to be cooled to
below the Fermi temperature \cite{jin99}.

An {\em a priori} candidate for the comparison between Bose and
Fermi gases is the simplest atom of all: hydrogen (H), with its
fermionic isotope deuterium (D). The appeal of these atomic gases
lies both in the possibility of cryogenic precooling, which
enables much larger samples to be cooled than with laser cooling
techniques, and in the existence of a natural two-dimensional
quantum gas on the surface of liquid helium. The H quantum gas has
been extensively investigated and quantum degenerate behaviour
(BEC) has been reported both in three dimensions \cite{fri98} and
in the two-dimensional (2-d) adsorbed gas \cite{saf98}. The first
optical observation of H atoms in the adsorbed phase was reported
in 1998\cite{mos98}. In that experiment Lyman-$\alpha$
spectroscopy at a wavelength of 122 nm was used to excite the
1S-2P transition in a surface specific way. In this paper we will
use Lyman-$\alpha$ spectroscopy to investigate a three dimensional
gas of atomic deuterium in contact with D atoms adsorbed on the
surface of liquid helium. Deuterium atoms are easily distinguished
from the hydrogen atoms that are always present as impurities,
because of the large isotope shift of about 630 GHz.

In contrast to H, very few experiments with spin polarized D have
been reported in the literature, mainly because spin-polarized D
has been found to be significantly less stable than spin-polarized
H and experiments are much harder
\cite{deut80,blue86,rey86,ara98}. The key reason for the low
stability of D is the high adsorption energy $E_a$ on the helium
surface which reduces the overall lifetime because loss processes
(molecule formation) occur primarily on the surface. Knowledge of
both this binding energy as well as the rate constant governing
the formation of molecules are essential ingredients for any
experiment aiming at 2-d quantum degeneracy. To date the values of
both these quantities are not beyond controversy. It has been
suggested \cite{ara98} that the recombination rate on the surface
in finite magnetic field may be dominated by a resonant process
similar to what was predicted for free D atoms \cite{rey88}. This
assumption has never been properly tested.

The aim of this paper is threefold: firstly, to provide a direct
measurement of the absorption energy of a deuteron onto liquid
helium, secondly to determine the rate constant for recombination
(molecule formation) of atoms residing on the surface as a
function of magnetic field, and finally to attempt to approach the
regime of Fermi-degeneracy in the adsorbate of D atoms as closely
as possible.

The adsorption energy for H on the surface of liquid helium
$E_a/k_{\rm B}=1.0$K \cite{blue86} is well established. For D the
values reported in literature vary between $E_a/k_{\rm B}=2.6(4)$
K \cite{deut80}, measured in 8T, and $E_a/k_{\rm B}=3.97(7)K$ in a
zero-field \cite{ara98}. The discrepancy between these values is
significantly larger than the accuracy of the measurements. A
possible explanation was put forward in ref. \cite{ara98}: It is
suggested that the presence of one or more Feshbach scattering
resonances may influence the measurement of $E_a$ when performed
in high magnetic field. In fact, measurements reported in ref.
\cite{deut80} and \cite{ara98} rely on the assumption that the
``cross length'' governing the formation of molecules on the
surface is independent of magnetic field and temperature. As we
will explain below, this assumption may be wrong if Feshbach
resonances are present.

All measurements of $E_a$ presented in literature for D atoms are
based on a determination of the decay the atomic gas due to
recombination of adsorbed atoms. A direct measurement of $E_a$ has
not been performed to date. Such a direct measurement relies on
the following simple relation, valid when a three-dimensional gas
is in equilibrium with an adsorbed two-dimensional phase in the
nondegenerate (Boltzmann) limit:
\begin{equation}
n_2=n_3 \Lambda \exp\left(\frac{E_a}{k_{\rm B}T}\right) \ ,
\label{adisotherm}
\end{equation}
where $n_2$ and $n_3$ are the surface and bulk density,
respectively, and $\Lambda=(2 \pi \hbar^2/m k_{\rm B} T)^{1/2}$ is
the thermal de Broglie wavelength. To obtain $E_a$ using this
equation one needs to have experimental access to $n_2$ and $n_3$
simultaneously. Using optical spectroscopy we were able to measure
$n_3$ and to obtain a fluorescence signal proportional to $n_2$
from the adsorbed atoms.

\section{Experimental methods}
\begin{figure}
\noindent \centerline{\epsfxsize=74mm\epsffile{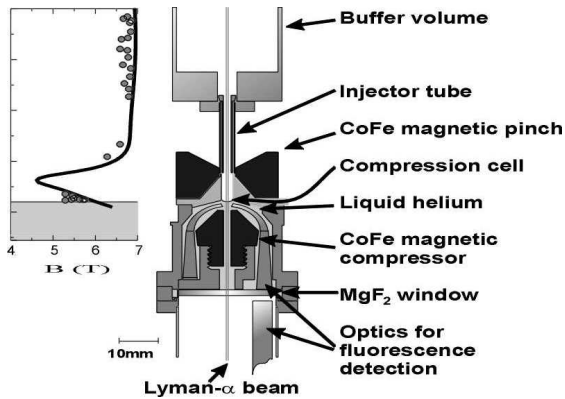}}
\vspace*{0.2cm} \caption{Drawing of the experimental apparatus.
The helium level and field profile are shown in the left panel.
The cell is placed in a 7 T bias field. The magnetic pinch and
compressor are cobalt-iron pole pieces producing a pronounced dip
(the magnetic barrier) and a local maximum (at the helium surface)
in the magnetic field. \label{fig:1}}
\end{figure}

We first describe the experimental setup used for the experiments
with H and D adsorbed on He. The heart of this setup is shown in
Fig.~\ref{fig:1}. A volume (called the buffer) of approximately 30
cm$^3$ is filled with D atoms, electron spin polarized into their
high-field seeking states. These atoms are produced in a cryogenic
discharge in helium vapor operated at a temperature of about 0.8K.
The discharge cell contains solid D$_2$ on the walls. These
molecules are dissociated while the helium discharge is running.
The produced monoatomic gas flows towards the buffer volume with a
flux of typically $10^{12}-10^{13}$ s$^{-1}$. We perform
experiments while the discharge is running with the filling flux
compensating the recombination losses. With the discharge switched
off the D gas decays in a few seconds due to the formation of
D$_2$ molecules.

\begin{figure}
\noindent \centerline{\epsfxsize=80mm\epsffile{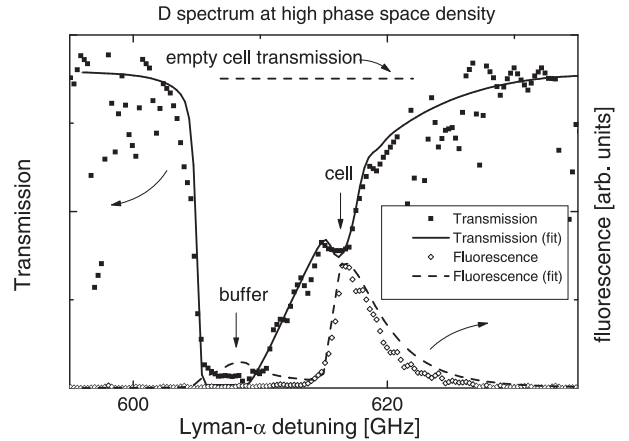}}
\vspace{0.2cm} \caption{LIF and absorption spectra for D atoms for
the transition between the levels $1S_{1/2}\ m_s=-1/2$ and
$2P_{3/2}\ m_j=-3/2$, taken with $\sigma_-$ polarized light. The
full symbols are data of the transmission through the cell and the
buffer. The solid line is a fit to the spectrum. The broad
saturated feature in the absorption spectrum is due to the buffer
atoms and the weaker satelite to the right is due to absorption by
the atoms in the cell. The open symbols are the fluorescence due
to bulk atoms in the cell. The signal of the surface atoms is not
visible in this plot. The noise in the absorption spectrum results
from falling He dropplets due to the fountain effect caused by the
temperature gradient between buffer and cell. The parameters
corresponding to the fits are: $n_{\rm b} \approx 6 \times
10^{11}$ cm$^{-3}$, $n_3 \approx 9 \times 10^{11}$ cm$^{-3}$. The
temperatures are $T_{\rm b}=0.44$ K,\ $T_{\rm c}=0.26$ K. The
detuning is measured relative to the center of the Lyman-$\alpha$
spectrum for H atoms. \label{fig:2}}
\end{figure}

Below the buffer volume there is a second volume called the cell.
The buffer is kept at a higher temperature than the cell to reduce
absorption of D on its walls. The atoms adsorb primarily on the
surface of a layer of bulk helium, located at the bottom of the
cell. The temperature of the liquid helium is about 0.3K. The
buffer temperature is typically between 0.4 and 0.45K. A strong
field gradient (1.6 T/cm) in the cell pushes the atoms towards the
helium surface, a principle known as magnetic compression. The
cell is separated from the buffer by a magnetic barrier (a local
field minimum) which limits the flux of atoms from the buffer to
the cell. The cell and buffer are mounted inside the bore of a
superconducting solenoid which produces a magnetic bias field of
between 4 and 7 T. The magnetic compression and the barrier
between buffer and cell are created with cobalt-iron pole pieces
(see Fig.~\ref{fig:1}).  We control the flux $\phi_{b\rightarrow
c}$ from the buffer to the cell by varying the temperature $T_b$
of the buffer volume. The flux is approximately given by:
\begin{equation}
\phi_{\rm b \rightarrow c} \approx \frac{1}{4} n_{\rm b} \bar{v}_b
A_{\rm b} \exp[-\mu_{\rm B}(B_{\rm b}-B_{\rm min})/k_{\rm B}T_{\rm
b}], \label{phibc}
\end{equation}
where $\bar{v}_b$ is the average thermal velocity of atoms in the
buffer and $\mu_{\rm B}$ is the Bohr magneton. $B_{{\rm b}}$ and
$B_{{\rm min}}$ are the magnetic fields in the center of the
buffer and at the magnetic barrier, respectively. $A_{\rm b}$ is
effective area of the aperture separating the buffer from the
cell, which includes the Clausing factor of the tube taking into
account the strong magnetic field inhomogeneity in the barrier
region. From a Monte Carlo calculation we obtained $A_{\rm
b}=1.57$ mm$^2$. The net flux from buffer to cell is $\phi_{\rm b
\rightarrow c} - \phi_{\rm c\rightarrow b}$ where the back flux
$\phi_{\rm c\rightarrow b}$ is given by the same expression
Eq.(\ref{phibc}) but with $n_{\rm b}$ replaced by the cell density
$n_3$, $B_{{\rm b}}$ replaced by the field $B_{{\rm c}}$ at the
bottom of the cell, and $T_{\rm b}$ replaced by the cell
temperature $T_{\rm c}$. The velocity $\bar{v}_c$ is evaluated at
the cell temperature $T_{\rm c}$ in this case.

A few words of clarification are needed about the meaning of the
cell density $n_3$. As in Eq.~(\ref{adisotherm}) we use the
subscript 3 (from 3-dimensional) for the cell density to
distinguish it from the 2-dimensional density of atoms adsorbed on
the surface. However, in contrast to the buffer where the field is
nearly constant, the cell has a field gradient and a concomitant
varying density. We will understand $n_3$ to mean the density just
above the surface. The density decreases with height above the
surface due to the strong field gradient, analogous to a
barometric height distribution. We now introduce a field and
temperature dependent effective volume $V_e=\int_V \exp(\mu_{\rm
B}[B({\bf x})-B(0)]/k_{\rm B}T) d^3x $. Here the integral is over
the volume $V$ of the cell and $B(0)$ is the field at the bottom
of the cell (i.e. at the helium surface). $V_e$ is defined such
that $n_3 V_e$ equals the number of atoms in the cell.

The atoms in the buffer, the cell, and the adsorbed phase, can all
be observed independently using Lyman-$\alpha$ spectroscopy. The
light source we used to produce tunable narrowband 122 nm light is
described elsewhere \cite{mos98,lui93}. As can be seen in
Fig.~\ref{fig:1} the light beam passes through the apparatus from
the bottom to the top. We measure both the absorption of the light
as well as the light-induced fluorescence (LIF) as a function of
the incident light frequency. The atoms in the buffer are
distinguished from those in the cell because they reside in a
different magnetic field and the resonance lines experience
different Zeeman-shifts. The LIF of adsorbed atoms can be
distinguished from that of those in the vapor because the
resonance lines are shifted by several hundred GHz relative to
those of bulk atoms due to the interactions with the liquid
helium.

\section{Experimental results}
In this section we discuss the experimental results of
spectroscopy of D atoms using Lyman-$\alpha$ spectroscopy. The
technique allows us to infer the three crucial parameters in the
problem: $n_b$,$n_3$, and $n_2$. The temperatures $T_b$ and $T_c$
are measured directly with resistance thermometers. In earlier
studies with H atoms \cite{mos98} it was found that the LIF of the
surface atoms takes the form of a broad feature shifted about
300GHz to the red of the spectral line of free atoms in the bulk
gas.
\begin{figure}
\noindent \centerline{\epsfxsize=80mm\epsffile{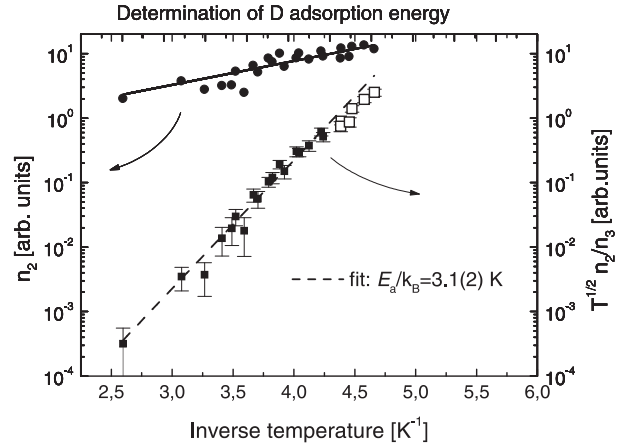}}
\vspace{0.2cm} \caption{Measured surface density in arbitrary
units (left scale) and ratio of cell density and surface density
multiplied by $\sqrt{T}$ in the cell. $n_2$ varies only weakly in
the temperature range shown, the buffer density decreases rapidly
with decreasing temperature. The solid points are used in the fit.
The open squares correspond to the temperature range were the 2-d
and 3-d systems become thermally decoupled due to fast surface
recombination. The fit gives a direct measurement of the binding
energy of 3.1 K. \label{fig:3}}
\end{figure}
In the deuterium case the broadening of the surface line is even
larger than for H. We have indication that the fluorescence is
broadenend to more than 1 THz. We could not detect a frequency
dependence over the 100 GHz lock range of our laser system and the
fluorescence was visible when a D atoms were present at all
frequencies between the H spectrum and a blue detuning of 100 GHz
with respect to the D lines. We infer the surface density of D
atoms by fixing the frequency of the light to a value close to
where the peak of the surface-LIF of the D atoms is expected,
about 400 GHz to the red of the free atom resonance. If there are
adsorbed atoms this gives us a LIF signal proportional to the
surface density. The vapor density above the surface $n_3$ can be
measured from the LIF as well as from absorption spectroscopy at
frequencies close to the free atom resonance. This is shown in
Fig.~\ref{fig:2}. The fit to the LIF signal of the bulk atoms is
based on a single-scattering model, which is a good approximation
at the relatively low densities involved. The model takes into
account the Zeeman and Doppler effects and the thermal
distribution of atoms in the cell. The shape of the curve is
additionally influenced by the solid angle to the detector. The
fluorescence detector can be calibrated in an absolute way by
comparing the fluorescence peaks with fits to the transmission
spectrum. From this fit we obtain an absolute value for $n_3$
whereas $n_2$ is determined up to a constant factor. We have made
these measurements over a range of temperatures, so that we can
use Eq.(\ref{adisotherm}) to directly obtain $E_a/k_B=3.1(2)$ K.
The result is given in Fig.~\ref{fig:3}. The range of temperatures
is limited from above by thermal instabilities (fountain effect)
of the helium in our apparatus, causing excess noise on the
transmission and fluorescence detectors. At the lowest
temperatures (for $1/T\agt4.5$K$^{-1}$), there is no thermal
equilibrium between the (quickly recombining) adsorbed atoms and
the bulk gas, and therefore Eq.(\ref{adisotherm}) is not
rigorously valid. We have discarded this temperature region from
the fit. In Fig.~\ref{fig:3}. we also plotted $n_2$ (in fact we
plot the fluorescence signal). It can be clearly seen that the
temperature dependence is quite different from that of the ration
$n_2/n_3$, which indicate that the fluorescence is not a spurious
artifact due to bulk atoms.

We can also use Eq.~(\ref{adisotherm}) to determine the maximally
obtained phase space density $n_2 \Lambda^2$, a measure of how far
we are from Fermi degeneracy. Although the measurement of $n_2$ is
not absolutely calibrated, the multiplicative factor follows
directly from Eq.~(\ref{adisotherm}). We find that at the lowest
temperature $T_c=0.19$K the phase space density is $\approx
10^{-2}$. In 2-d this implies $T_F/T \approx 10^{-2}$ where $T_F$
is the Fermi temperature.

The value of $E_a/k_{\rm B}=3.1(2)$K obtained from the fit in
Fig.~\ref{fig:3} is consistent with the 2.6(4) K measured by
Silvera and Walraven \cite{deut80} but lower than the value
3.97(7) K reported by Arai et al. \cite{ara98}. In the latter
paper it is suggested that the difference may be explained by the
fact that their experiment is performed in zero field whereas
Silvera and Walraven use a field of 8T. The way in which $E_a$ is
measured in both these experiments is by determining the decay
rate of the gas. At sufficiently low temperature this decay takes
place almost exclusively by formation of molecules from adsorbed
atoms on the surface. Consequently the natural way to describe the
population dynamics would be in terms of an inelasic cross length
$l_{DD}$ governing the collisions between to adsorbed atoms which
lead formation of a D$_2$ molecule. The rate of the process is
proportional to $n_2$ and to the effective surface area $A_2$ on
which the binary collisions take place. This surface is defined
analogous to $V_e$ as: $A_2=\int_{{\rm surface}} \exp(2\mu_{\rm
B}[B({\bf x})-B(0)]/k_{\rm B}T) d^2x$, where the integral runs
over the surface of the cell. The factor 2 appears in the exponent
because two atoms are involved in the formation of molecules. Both
$V_e$ and $A_2$ depend on temperature. Our recombination
measurements are performed at $T_c=0.3$ K. At this temperature we
have for our cell: $V_e=0.015$cm$^3$ and $A_2=0.14$cm$^2$.

The surface density is normally difficult to measure and as a
consequence in the literature the decay is usually described in
terms of an {\em effective} bulk recombination rate constant $K$.
If we assume that the gas has no nuclear polarization (equal
population of all high-field seeking hyperfine states) we may
write: \cite{rey86,ara98}:
\begin{equation}
\dot{N}=-V_{e}K n_3^2, \label{Ndot}
\end{equation}
where $\dot{N}$ is the decay of the 3-d gas in the cell. The
effective two body decay rate constant $K$ is given by:
\begin{equation}
K=l_{\rm DD}\bar{v}\frac{A_2}{V_{e}}\Lambda^2
\exp\left(\frac{2E_a}{k_{\rm B}T}\right).
\label{K}
\end{equation}

\begin{figure}
\noindent \centerline{\epsfxsize=80mm\epsffile{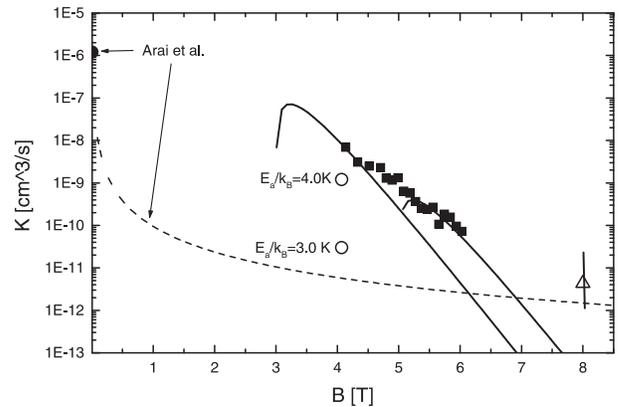}}
\vspace{0.2cm} \caption{Rate constant for 2 body recombination
(full symbols) at for $T_{\rm b}=0.43$K and $T_{\rm c}=0.30$K,
compared to results from ref. 7 (triangle) and 9 (circle). The
error margin on the result from ref. 7 is the result of the
exrapolation to our temperature. The solid circle and the dotted
line are the measurement of ref. 11 at zero field and the
extrapolation to finite fields, respectively. The solid curves are
resonance curves taken from ref. 10, scaled to appropriate
strength and shifted to 3 and 5 T respectively. \label{fig:4}}
\end{figure}

Here $\bar{v}$ is the average thermal velocity of adsorbed atoms.
Eqs.(\ref{Ndot}) and (\ref{K}) are based on the physical picture
that atoms spend part of their time on the surface and that in
equilibrium the fraction of particles on the surface is determined
by Eq.(\ref{adisotherm}). In the exponent a factor $2E_a$ appears
because molecules are formed in pair collisions, hence the process
depends on $n_2^2$. By measuring $\dot{N}$ and $n_3$ as a function
of $T$ one in principle obtains values for $E_a$ and $K$. If one
assumes that $l_{\rm DD}$ is independent of $B$ and $T$ one
obtains the following relation between the $K$ in zero and finite
fields \cite{ara98}:
\begin{equation}
K(B)=\frac{8}{3}K(0) \varepsilon^2_D, \label{extrapolation}
\end{equation}
where $\varepsilon_D= a_D/ (\sqrt{2}\mu_B B)$ with $a_D/h=218$ MHz
the hyperfine frequency.

The way just described to determining $E_a$ is plagued by the
uncertainty weather $l_{\rm DD}$ is in fact independent of $B$ and
$T$. Therefore we have independently measured $E_a$ and $l_{\rm
DD}$. We have measured the field dependence of $K$ in the
following manner: The discharge is kept running, producing an
approximately constant flux of D atoms. Some of the atoms
recombine in the buffer volume but most escape over the barrier to
the cell. The buffer density is measured by fitting the absorption
spectrum of the buffer atoms. From this we calculate the escape
flux $\phi_{\rm b \rightarrow c}$ from buffer to cell according
Eq.(\ref{phibc}). From the fluorescence spectrum of the cell we
calculate $n_3$ which gives us $\phi_{\rm c\rightarrow b}$, which
is generally much smaller than $\phi_{\rm b \rightarrow c}$. In
steady state the quantity $\dot{N}$ in Eq.~(\ref{Ndot}) is equal
to the net flux into the cell $\phi_{\rm b\rightarrow c}-\phi_{\rm
c\rightarrow b}$. From this, with Eq.~(\ref{Ndot}) we obtain the
rate constant $K$. The result is plotted in Fig.~\ref{fig:4} as a
function of $B$.

Since the result may be thought to depend exponentially on
thermometry errors, we took spectra of the H which occurs as an
impurity in our deuterium, simultaneously. Since the H does not
recombine quickly, it can be used to check the balance of the
fluxes:  for H the flux into the cell must equal the return flux.
We found the ratio of the density of H atoms in the buffer and the
cell to be within the $30\%$ accuracy of the measurement. This
indicates that possible thermometry errors (due to e.g. spurious
thermometer magnetoresistance) cannot contribute more than a
factor $\approx 2$ error to the measured field dependence while
the data cover several orders of magnitude over the measures field
range.

\section{Discussion}
In Fig.~\ref{fig:4} we compare our data to values obtained by
other groups. The dotted line in the figure is an extrapolation to
finite magnetic field of the zero field measurement of Arai et al.
\cite{ara98}. We have scaled their zero field measurement to
correspond to our temperature (the cross in Fig.~\ref{fig:4}). To
extrapolate their measurement to finite fields we used
Eq.~(\ref{extrapolation}). All data compiled in Fig.~\ref{fig:4}
have been rescaled to correspond to our value of
$A_2/V_e=9$cm$^{-1}$. It is clear that our data lie well above the
curve of Arai et al. Moreover, there is a much stronger dependence
on magnetic field. This can not be due a spurious effect of
electron-spin relaxation as for our field range and temperature
this lies many orders of magnitude below our measured range of $K$
values. It was suggested in ref.\cite{ara98}, on the basis on an
analogy with the situation for free D atoms \cite{rey88}, that $K$
may be enhanced in finite field due to Feshbach resonances. Such a
resonance occurs when the energy of a highly excited state of the
singlet molecule coincides with the energy of triplet pair of
deuterons in the high field seeking state; the two channels are
then coupled by the hyperfine interaction. Near such a resonance
we may write $K$ as:
\begin{equation}
K_{\rm res}=\frac{A_2}{V_{e}} \frac{\Lambda^4}{\tau}
\exp\left(\frac{2E_a-2\mu_{\rm B}B+E_{\rm D}}{k_{\rm B} T}\right).
\label{Kres}
\end{equation}
Here $E_{\rm D}$ denotes the dissociation energy of a singlet
molecule into a pair of adsorbed atoms. Expression (\ref{Kres}) is
valid for fields higher than $E_{\rm D}/2\mu_{\rm B}$. The
reaction time $\tau$ characterizes a two-step process: the
formation of the excited singlet state molecule followed by its
decay to lower lying levels.

For free D atoms two Feshbach resonances, 2 T apart, in which the
rovibrational levels $(v,J)=(21,0)$ and $(21,1)$ of $D_2$ are
involved, were predicted \cite{rey88} in field range relevant for
this experiment. This suggests that we compare our data with with
curves corresponding to Eq.(\ref{Kres}). Two such curves are
plotted in Fig.~\ref{fig:4} at the somewhat arbitrarily chosen
field values of 3 and 5 T (with the vertical scale adjusted to
match the data best). Above of a single resonance the rate will
vary proportional to $\exp(-g \mu_B B/k_B T)$. The data vary
slightly more smoothly with $B$. It is to be expected that the
free atom resonances are broadened in the presence of a helium
substrate. One might expect a smearing of the same order of
magnitude as $E_a$, which would correspond to a width of more than
4 T, enough to wash out any structure in the resonance spectrum.
Another possibility is that motion of the singlet molecule coming
off the surface engenders a continuous spectrum of resonances for
surface recombination. At any rate the strong observed field
dependence supports the idea of resonance recombination.

Three other data points are included in Fig.~\ref{fig:4}. The
measurement at 8 T taken from ref.\cite{deut80}, extrapolated to
$T=0.3$K using their measured temperature dependence and corrected
for our value of $A_2/V_e$, is consistent with the extrapolation
of our data to higher field. The error results from the
uncertainty in the value for $E_a$ quoted in ref.\cite{deut80}
which is reflected in the extrapolation to lower temperature. In
fact our data indicate that the influence of the resonance hardly
extends to a field of 8 T which explains why the value for
$E_a/k_B=2.6(4)$K found in ref.\cite{deut80} coincides with ours
within the margin of error. The datapoints at 4.07T are taken from
ref.\cite{rey86}. We have extrapolated the measurements taken at
their lowest temperature of 0.65K down to our cell temperture
using two different values: 3.0 and 4.0 K, respectively, for the
effective adsorption energy (asuming $K\sim T^{-1/2}
\exp[2E_a/k_BT]$ for the extrapolation). Both datapoints obtained
in this way from ref.\cite{rey86} lie below ours. The reason for
this discrepancy is unclear.

In conclusion, the strong field dependence of the recombination
cross length of deuterium revealed by our measurements does point
towards a resonance recombination process on the surface. In
higher field the recombination rate rapidly becomes smaller. Our
results indicate that at fields above the resonant region (roughly
$B>8$T) the recombination cross lengths may become sufficiently
rather small to make adsorbed D nearly as stable as adsorbed H.
Hence one might hope that the interesting configuration of a
degenerate, two-dimensional Fermi gas on the surface of a Bose
liquid is attainable in this way.

We are grateful to Pavel Bushev, Tycho Sonnemans, and Jook
Walraven for fruitful discussions and comments. This research is
supported by the Stichting Fundamenteel Onderzoek van der Materie
(FOM).

\end{document}